\documentclass[prd,aps,floats,floatfix,superscriptaddress,preprintnumbers,showpacs,eqsecnum,nofootinbib,twocolumn]{revtex4}
\usepackage{latexsym,array,theorem,mathrsfs,bm,float}

\usepackage{psfrag}
\usepackage{amsfonts,amsmath,amssymb,latexsym,array,afterpage,theorem,mathrsfs,bm,float,epsfig,color,graphicx,tabularx,here,multirow}

\definecolor{red}{rgb}{1,0,0}
\definecolor{blue }{rgb}{0,0,1}
\definecolor{green}{rgb}{0,1,0}


\newcommand{\bea}{\begin{eqnarray}}
\newcommand{\ena}{\end{eqnarray}}
\newcommand{\beann}{\begin{eqnarray*}}
\newcommand{\enann}{\end{eqnarray*}}

\newcommand{\dsl}{\pa \kern-0.5em /}

\newcommand{\pa}{\partial}

\newcommand{\nn}{\nonumber\\}

\newcommand{\vect}[1]{\!\!\!\mbox{\,~\boldmath $#1$}}

\usepackage{graphicx}
\usepackage{dcolumn}
\usepackage{amssymb}
\usepackage{pifont}

\usepackage{bm}

\newcommand{\be}{\begin{equation}}
\newcommand{\ee}{\end{equation}}

\newcommand{\mpl}{M_\mathrm{\rm P}}

\begin{document}

\baselineskip=12pt

\preprint{WU-AP/1706/17}

\title{Hybrid Higgs Inflation: The Use of Disformal Transformation}

\author{Seiga Sato}
\email{s.seiga"at"gravity.phys.waseda.ac.jp}
\author{Kei-ichi Maeda}
\email{maeda"at"waseda.jp}
\affiliation{
Department of Physics, Waseda University,
Shinjuku, Tokyo 169-8555, Japan}

\date{\today}

\begin{abstract}
We propose a hybrid type of the conventional Higgs inflation and new Higgs inflation models.
We perform a disformal transformation into the Einstein frame and analyze the background dynamics and the 
cosmological perturbations in the truncated model, in which we ignore the higher-derivative terms of the Higgs field.
From the observed power spectrum of the density perturbations, 
we obtain the constraint on  the non-minimal coupling constant $\xi$ and the mass parameter $M$
 in the derivative coupling. 
Although the primordial tilt $n_s$ in the hybrid model barely changes,  the tensor-to-scalar ratio $r$ 
moves from  the value in new Higgs inflationary model to that in the conventional Higgs inflationary model as 
 $|\xi|$ increases.
We confirm our results by numerical analysis by ADM formalism of the full theory in the Jordan frame.
\end{abstract}

\maketitle




\section{Introduction} 

Inflation was first proposed to solve the flatness and  the horizon problem as well as the monopole problem
in the early universe\cite{Starobinsky:1980te,Guth:1980zm,Sato:1980yn}.  It also predict the origin of cosmological perturbations
\cite{Hawking:1982cz,Guth:1982ec,Starobinsky:1982ee,Mukhanov:1982nu}, which turns out to be the most 
important outcome for the observational confirmation of the idea.
 Since the generic predictions of the inflationary scenario has
matched with observations, inflation has now become a standard model of the early universe.

Many inflation models have  been so far proposed\cite{Martin:2013tda}, but 
 the constraints from the recent CMB observations are very severe.
The density perturbations are almost scale invariant, while 
the tensor-to-scalar ratio should be very small.
As a result, some inflation models have been already excluded.

In most inflationary scenarios, we assume the existence of a scalar field which is responsible for inflation,
which is called an inflaton. 
However we still do not know what is the inflaton.
Since we know only one scalar field  in the standard model, which is the Higgs particle,  
many models have looked for an inflaton beyond the standard model of particle physics.
It is because the Higgs field cannot be an inflaton to explain the observed density perturbations.

However there is one loophole,
 which is an introduction of gravitational couplings with the Higgs field.
When we quantize matter fields in a curved space time, 
we may find non-minimal coupling of the fields with a curvature.
Including such a coupling term $\xi \phi^2 R$, where $\xi$ is a non-minimal coupling constant and $R$ is a scalar curvature,
  some inflationary scenario has been discussed
\cite{spokoiny1984inflation, Maeda:1988rb, futamase1989chaotic}.
Assuming  this scalar field is the Higgs particle, 
which is only one scalar field in the standard model, 
 the so-called  Higgs inflation has been proposed
 \cite{Bezrukov:2007ep,Bezrukov:2008ej,DeSimone:2008ei,Bezrukov:2009db,Bezrukov:2013fka}.
 We call it the conventional Higgs inflation model.
If $\xi$ is large negative, e.g.,  $\xi\sim-10^4$, this model 
provides a consistent inflationary scenario with observations.

New Higgs inflation was also proposed in 2010, which has a derivative coupling of 
the Higgs field with the Einstein curvature as ${1\over 2M^2}G^{\mu\nu}\nabla_\mu \phi\nabla_\nu\phi$
 \cite{Germani:2010gm,Germani:2010ux}.
When $M\sim10^{-7} \mpl$, where $\mpl$ is the reduced Planck mass, 
this model is still consistent with observations.
Although this model produces a rather large amount of the gravitational waves,  
the predicted tensor-to-scalar ratio is still just on the border of the observational constraints.

The common feature  of both Higgs inflation models is no introduction of any 
additional degrees of freedom.
Although we still do not know the physics at the very high energy scale around the Planck scale ($\sim10^{18} $ GeV),
the above coupling terms may appear at the high energy scale\cite{kamada2012generalized}.
Hence, it may be natural to consider a hybrid type of the above two inflation models.

Since the typical energy scale of inflation is very high, 
we have to take into account the quantum loop effects,
which may destabilize the Higgs field\cite{Degrassi:2012ry,Arnold:1989cb,Anderson:1990aa,Arnold:1991cv,Espinosa:1995se,Buttazzo:2013uya,EliasMiro:2011aa,Shaposhnikov:2014aga,Espinosa:2015qea,bezrukov2015living,Rose:2015lna}. 
Whether 
it is really unstable or stable may depend highly on the top quark mass.
Even if it is stable,  since the running coupling constants via renormalization 
depend on the energy scale, the effective potential will be modified much\cite{Sher:1988mj,Hamada:2014iga,Bezrukov:2009db,DeSimone:2008ei,Barvinsky:1993zg,Barvinsky:2009ii,Abe:2016irv,Smolin:1979ca,George:2013iia,George:2015nza} .
Although we may have to analyze our model with the loop corrections, 
since the discussion on the loop corrections is still in a fog, 
we will analyze the original tree model in this paper
 to show how the tensor-to-scalar ratio depends on the coupling parameters.

This paper is organized as follows.
After introduction of the hybrid type Higgs inflation, 
we present our truncated model based on a disformal transformation
into the Einstein frame 
and analyze the background cosmological dynamics
in \S.II.  We confirm the validity of our truncation by 
the results calculated in the original full theory.
In  \S.III,  we analyze the cosmological perturbations 
and give the primordial tilt $n_s$ and the tensor-to-scalar ratio $r$.
We then show  the accuracies of our results obtained in the truncated model,
which confirm our approach.
Summary and discussion follow in \S.IV.
In Appendix A, we present the explicit forms of the disformal transformation
and the higher-derivative terms of the Higgs field.
We then give the basic equations for the background universe  in the original Jordan frame in Appendix B.
We also describe how to calculate  the cosmological perturbations in the full theory by use of 
ADM formalism.

In this paper we use the reduced Planck units of $\mpl:=(8\pi G)^{-1/2}=1$ as well as $\hbar=c=1$.

\section{Hybrid Higgs Inflation}
First we summarize the original two Higgs inflation models and then extend them to a hybrid type.

\subsection{Two original Higgs inflation models}

The action of the conventional Higgs inflation based on non-minimal coupling of the Higgs field with a scalar curvature
\cite{Bezrukov:2007ep} is given by
\beann
S_J&=&\int d^4x\sqrt{-g}\Big[ \frac{1-\xi \phi ^2}{2}  R(g)
\\
&&~~~~~~~~~~~~~~~-{1\over 2}\left(\nabla \phi\right)^2-V(\phi) \Big]
\,,
\enann
where  the potential of the Higgs field $\phi$ 
is given by 
\beann
V(\phi) =
\frac{\lambda}{4}(\phi^2-v^2)^2
\,.
\enann
 $\lambda$ is the Higgs self-coupling and $v (\approx 246\text{ GeV})$ is the Higgs vacuum expectation value (VEV)
\cite{Bezrukov:2007ep}. 
When we discuss inflation, which energy scale may be much higher than 
100 GeV, we can approximate the potential as
\beann
V(\phi) =
\frac{\lambda}{4}\phi^4
\,.
\enann
For the successful Higgs inflation, we assume $\xi<0$.
Note that the sign of the non-minimal coupling $\xi$ is different from
that in \cite{Bezrukov:2007ep}.

Performing a conformal transformation
\beann
g_{\mu\nu}=\left(1-\xi\phi^2\right)^{-1}\bar g_{\mu\nu}
\,,
\enann
we find the Einstein frame action
\beann
	S_E
	=
	\int d^4x \sqrt{-\bar g}  \left[
	{1\over 2} R(\bar g)
		- {1\over 2}(\bar \nabla \Phi)^2
		- U(\Phi)
	\right]
\,,
	\label{eqn:conv_higgs_action}
\enann
where the variables with  a bar denote those in the Einstein frame.
 $\Phi$ is the canonically normalized Higgs field, defined by 
\beann
{d\Phi\over d\phi}={\sqrt{1-\xi(1-6\xi)\phi^2}\over (1-\xi \phi^2)}
\,,
\label{Phiphi}
\enann
 and 
\beann
	U(\Phi)
	=
	\frac{\lambda \phi^4}{4\left( 1-\xi \phi^2\right)^{2}}
	\label{eqn:conv_higgs_potential}
\enann
is the potential in the Einstein frame. 

When $|\xi|\gg 1$, we find
 \beann
 \Phi\approx \sqrt{3\over 2} \ln 
 \left(1-{\xi\phi^2}\right)
 \,,
 \enann
which gives 
 \beann
	U(\Phi)
	&\approx  &
	\frac{\lambda }{4\xi^2}
	\left( 1-e^{-\sqrt{2\over 3}
	{\Phi}}\right)^{2}
	\,.
	\label{eqn:conv_higgs_potential2}
	\enann
This model is almost equivalent to the Starobinsky inflation with the curvature squared term\cite{Starobinsky:1980te}.
When we perform the conformal transformation\cite{Maeda:1987xf, Maeda:1988ab}, 
the model is described by the Einstein gravity plus a scalar field with the same potential..

This potential approaches a constant 
exponentially as $\Phi\rightarrow \infty $, 
which guarantees 
de Sitter expansion.  
Assuming $\lambda \simeq 0.1$, if $\xi\sim -10^{4}$, we obtain
 the cosmological perturbations consistent with observations as 
the primordial tilt $n_s\sim 0.96$ and
 the tensor-to-scalar ratio
$r\sim 10^{-5}$ as well as the power spectrum of the density perturbations ${\cal P}_\zeta\sim 10^{-9}$\cite{Bezrukov:2007ep} .

An alternative approach, called new Higgs inflation, adds a coupling between derivatives of the Higgs field with the Einstein curvature, 
which action is given by 
\beann
	S_J
	&=&
	\int d^4x \sqrt{-g}  \bigg[
\frac{1}{2} R(g)
\\
&&
-{1\over 2}\left( g^{\mu\nu}-\frac{G^{\mu\nu}}{M^2}\right) \nabla_\mu \phi \nabla_\nu \phi-V(\phi) 	
	\biggr]
   \,.
\enann
where $G^{\mu\nu}$ is the Einstein tensor
and $M$ is a coupling constant described by a characteristic mass scale
\cite{Germani:2010gm,Germani:2010ux,Bezrukov:2013fka}. 
Like the conventional
Higgs inflation, this does not introduce extra degrees of freedom.  
The new coupling constant
takes on a value $M \simeq 5 \times 10^{-8} \lambda^{-1/4}$
when normalized with CMB data.  We find $n_s \sim 0.97$ and $r\sim 0.1$\cite{Germani:2014hqa} .


\subsection{Hybrid Model}

Here we propose a hybrid type of the previous two Higgs inflation models\cite{hybrid_JPS}.
The action we assume is
\bea
S_J&=&\int d^4x\sqrt{-g}\Big[ \frac{1-\xi \phi ^2}{2} R(g)
\nn
&&
-{1\over 2}\left( g^{\mu\nu}-\frac{G^{\mu\nu}}{M^2}\right) \nabla_\mu \phi \nabla_\nu \phi
-V(\phi) \Big]\,,~~~~~
\label{hha}
\ena
 which we call a hybrid Higgs inflation model.

This type of model is classified into more general 
model proposed 
 in \cite{kamada2012generalized}, which they called generalized Higgs inflation.
They discussed the model in the original Jordan frame.
Since the basic equations are complicated, some limiting case was discussed to 
present a simple formula for the cosmological perturbations.
However it turns out that  their approximation cannot be applied to  the conventional Higgs inflation model
(see later). 

Hence in this paper we shall perform a disformal transformation to the Einstein frame 
and ignore the higher-derivative terms\cite{di2016electroweak}, 
which may be justified 
during the slow-rolling inflationary period as we will show the detail later.
We show how to get the Einstein frame via a disformal transformation 
 in Appendix \ref{Towards_Einstein}.

\begin{figure}[h]
 \begin{center}
 \includegraphics[width=7cm]{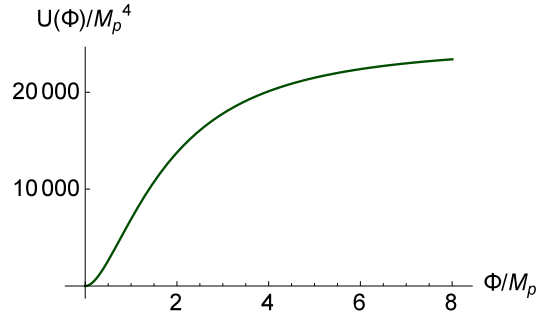}
 \end{center}
 \caption{The effective potential $U(\Phi)$ in the Einstein frame for the coupling parameters
 $M=10^{-7} , \xi=-10^{4}$ and $\lambda=0.1$. 
 }
 \label{U-phi}
\end{figure}

The action (\ref{hha}) in the Einstein frame is described as
\bea
S_E&=& \int d^4x\sqrt{-\bar{g}}\Big[\frac{1}{2} \bar{R}-{1\over 2}\left(\bar\nabla\Phi\right)^2-U(\Phi )
\nonumber 
\\
&&+{\rm higher~derivative~terms}\Big]
\,,
\label{dhha}
\ena
 where the variables with  a bar denote those in the Einstein frame. 
New canonically normalized scalar field $\Phi$ is
defined by 
\beann
\frac{d\Phi}{d\phi}=\left(1-{\xi\phi^2}\right)^{-1}\sqrt{1-{\xi(1-6\xi)  \phi^2}+\frac{V}{M^2}}.
\label{cns}
~~
\enann
The  effective potential in the Einstein frame 
$U(\Phi)$ is given by 
\beann
U(\Phi)={V\left(\phi\right)\over \left(1-{\xi \phi^2}\right)^{2}}
\,,
\enann
in the form of  the parametric representation by the original Higgs field $\phi$.
Fig. \ref{U-phi} shows the schematic shape of $U(\Phi)$.

In the transformed action (\ref{dhha}), there are many higher derivative terms such as $(\nabla\phi)^4$ and $\Box \phi$. 
In Appendix \ref{Towards_Einstein}, we present  the higher-derivative terms explicitly.
However, since we discuss the inflationary period, in which
the inflaton changes slowly, we may be able to ignore those terms.
The resultant truncated action is now 
\bea
S_{E}^{\rm (T)}=\int d^4x\sqrt{-\bar{g}}\left[\frac{1}{2} \bar{R}-{1\over 2}\left(\bar\nabla\Phi\right)^2
-U(\Phi )\right]
\,.
\nn
\label{dthha}
\ena
If this action approximates the original one well, 
we can easily analyze the inflationary stage just by the potential $U(\Phi)$.
This is  the great  advantage of the use of a disformal transformation.
This potential is expanded in the large field limit as
\beann
U(\Phi)&=&
{\lambda\over 4\xi^2}\Big{[}1-{\lambda  \over 
2|\xi|^3 M^2 }
 {1\over \Phi^2}+\cdots 
 \Big{]}\,.
 \label{eqn:new_potential}
\enann
This inflation potential shape is favored by the CMB observation because it is concave.
This potential asymptotically converges to be a flat plateau in the large field region as $\Phi^{-2}$ 
unlike the conventional Higgs inflation, which approaches 
a constant exponentially as $\exp\left(- 2\sqrt{2\over 3}\Phi\right)$.
 
 In what follows, we analyze this truncated model and then justify this approach by comparing 
with the results obtained  without truncation in the original Jordan frame.

\subsection{Background Dynamics}
\label{background_dynamics}

In this subsection, we first analyze an isotropic and homogeneous universe model.
We assume that the metric form is given by the flat FLRW (Friedmann-Lem\^{a}itre-Robertson-Walker)  spacetime as
\beann
d\bar s^2=-d\bar t^2+\bar a^2 d\bar{\vect{x}}^2
\,.
\enann

When we ignore the higher-derivative terms, the basic equations in the Einstein frame are given by 
\begin{align}
&\bar{H}^2=\frac{1}{3}\left[\frac{1}{2}\left({d\Phi\over d\bar t}\right)^2+U(\Phi) \right]
\label{dfr1}\\
&\dot{\bar{H}}+\bar{H}^2=-\frac{1}{3}\left[\left({d\Phi\over d\bar t}\right)^2-U(\Phi) \right]
\label{dfr2}\\
&{d^2\Phi\over d\bar t^2}+3\bar{H}{d\Phi\over d\bar t}+\frac{dU}{d\Phi}=0
\,,
\label{deom}
\end{align}
where the Hubble expansion parameter is defined by 
\beann
\bar H={1\over \bar a}{d\bar a\over d\bar t }
\,.
\enann

Since the potential $U(\Phi)$ in the Einstein frame has a very flat plateau at the large field range,
 we will have a slow-roll inflation. 
Once the scalar field starts to roll the potential very slowly, the higher-derivative terms can be ignored.
As a result, we expect the truncated model (\ref{dthha}) may approximate an inflationary 
period very well.

In order to justify this expectation, we solve 
both cosmological dynamics in the original model and
  in the truncated model, and compare those results.
  Since the full equations in the Einstein frame are too complicated, we solve 
  them in the original Jordan frame, which basic equations are given in Appendix \ref{basic_equations_Jordan}.
Eqs. (\ref{fr1}),(\ref{fr2}) and (\ref{eom}) in the original 
Jordan frame correspond to Eqs (\ref{dfr1}),(\ref{dfr2}) and (\ref{deom}), respectively.

Note that the cosmic time $\bar t$ in the Einstein frame is different from the cosmic time $t$ in the Jordan frame.
Since two metrics are related as 
\beann
ds^2&=&g_{\mu\nu}dx^\mu dx^\nu=-d t^2+ a^2 d\vect{x}^2
\\
&=&\left(1-{\xi\phi^2}\right)^{-1}\left[
d\bar s^2+{\nabla_\mu\phi\nabla_\nu\phi\over 2M^2}dx^\mu dx^\nu\right]
\\
&=&\left(1-{\xi\phi^2}\right)^{-1}\left[
-d \bar t^2+\bar a^2 d\vect{\bar x}^2+{\dot \phi^2 \over 2M^2}dt^2\right]
\,,
\enann
where a dot denotes the derivative with respect to  the cosmic time $t$ in the Jordan frame,
we find the relations between two cosmic times ($t$ and $\bar t$) and two scale factors 
($a$ and $\bar a$) as
\beann
&&
d\bar t^2=\left(1-{\xi\phi^2}+
{\dot \phi^2 \over 2M^2}\right)\, dt^2
\\
&&
\bar a^2=\left(1-{\xi\phi^2}\right)\, a^2
\,,
\enann
respectively.

\begin{figure}[h]
 \includegraphics[width=8cm]{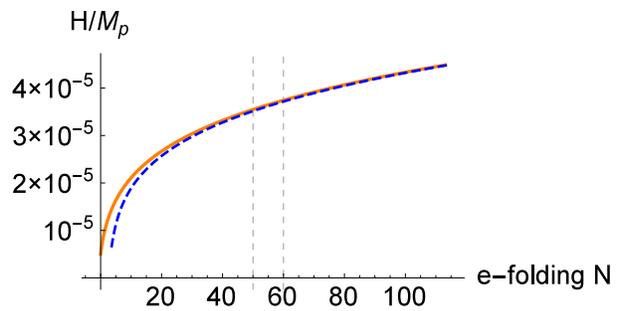}
 \caption{The background dynamics for the coupling parameters $M=10^{-7},\xi=-10^4$. 
 We show the evolution of the Hubble expansion parameter $H$ with respect to the e-folding number $N$ in the Jordan frame.
 The orange curve denotes the evolution of the full theory in the Jordan frame, while
  the blue dotted curve shows the results obtained by the truncated model in the Einstein frame.
  We transform its solution into the Jordan frame variables.
We also show the reference e-folding number of $N=50-60$ by the dotted light-gray lines.}
 \label{hubble}
       

\end{figure}

We numerically solve the basic equations, and compare both dynamics.
Fig. \ref{hubble} shows the time evolution of the Hubble expansion 
parameter $H$ in the Jordan frame. 
In stead of the cosmic time to describe the evolution,
we use the e-folding number $N$ to the end of inflation, which is defined by
\beann
N=\ln \left({a_{\rm end }\over a}\right)
\,,
\enann
where $a_{\rm end}$ is the scale factor at the end of inflation.
In order to compare the numerical solutions, 
the solution of the truncated model in the Einstein frame 
is transformed to the Jordan-frame variables.

Although there exists some small deviation at the end of inflation ($N=0$),
when the scalar field starts to move rapidly, the truncated model approximates the original one very
well until $N\approx 20$.  Especially around $N=50-60$, when the observed density perturbations are produced,
the agreement is quite good.

\begin{figure}[h]
\includegraphics[width=8cm]{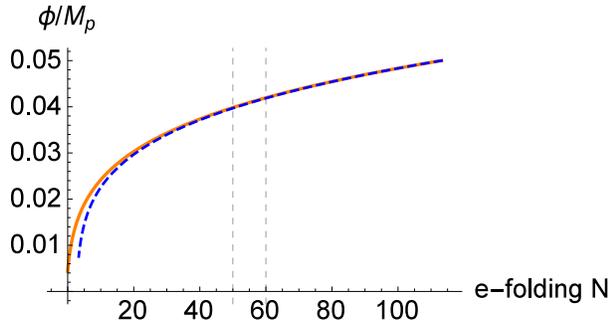}
 \caption{The evolution of the Higgs field $\phi$ with respect to the e-folding number $N$.
The coupling parameters, the colored curves, and the dotted  light-gray  lines  are the same setting as
those  in Fig. \ref{hubble}} 
 \label{infla}

\end{figure}

\begin{figure}[h]
 \begin{center}
 \includegraphics[width=8cm]{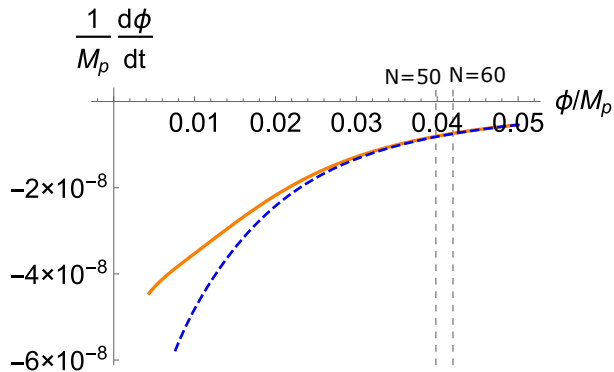}
 \end{center}
 \caption{The phase-space  evolution of the Higgs field ($\phi$-$\dot \phi$).
The coupling parameters, the colored curves, and the dotted light-gray lines  are the same setting as
those  in Fig. \ref{hubble}}
 \label{phase}
\end{figure}

We also depict the evolution of the Higgs field $\phi$ and
 its phase diagram (the $\phi$-$\dot \phi$ diagram) in
Fig. \ref{infla} and
Fig \ref{phase}, respectively.
These also support the validity of our truncation.
In the early stage of inflation,
both background solutions are almost the same.
At the end of inflation, the deviation appears since 
the higher-derivative terms seem no longer negligible.


In the next  section,
we will discuss the cosmological perturbations  and the validity of the truncated model.

~~

\section{Cosmological perturbations}
\label{cosmological_perturbations}

\subsection{Perturbations in the truncated model}
\label{perturbations_truncated}

In order to confirm our inflationary model, we have to evaluate the cosmological perturbations and compare them with the observed data.
As shown in \cite{makino1991density, minamitsuji2014disformal, tsujikawa2015disformal, watanabe2015multi, domenech2015cosmological, motohashi2016disformal}, a disformal transformation as well as a conformal transformation  
will not change the cosmological perturbations.  There exists the invariance of the cosmological perturbations under such transformations.
Hence, we can evaluate the perturbations either in the Einstein frame 
or in the Jordan frame.
Since the full equations in the Einstein frame are very complicated, we 
usually analyze the dynamics in the original Jordan frame.
In fact, in generalized Higgs inflation model, which includes ours, 
they have analyzed the perturbations in the Jordan frame\cite{kamada2012generalized}.
Since the full analysis of perturbations is tedious, 
they have also provided one simple formula under some approximation
\cite{kamada2012generalized}.

However, as shown in the subsection \ref{background_dynamics},
during the inflationary stage when the scalar field changes very slowly,
the background dynamics in the truncated model well approximates 
that in the original full model. 
Hence we expect that the cosmological perturbations in the truncated model
may also provide a good approximation to the original ones.

Here we first evaluate  the cosmological perturbations in the truncated model  in the Einstein frame, in which the perturbations can be obtained easily 
just by the effective potential $U(\Phi)$. 
This is a great advantage of the present method.

The potential slow-roll parameters are defined by
\beann
\epsilon&=&\frac{1}{2}\left(\frac{1}{U}\frac{dU}{d\Phi} \right)^2\,,
\\
\eta&=&\frac{1}{U}\frac{d^2U}{d\Phi^2}.
\enann
The power spectrum of the density perturbations $\mathcal{P}_\zeta$,
its spectral index  $n_s$ and the tensor-scalar ratio $r$ are given by the slow-roll parameters as
\beann
\mathcal{P}_\zeta &\simeq& \frac{1}{24\pi^2}\frac{U}{\epsilon},
\label{psp}\\
n_s&\simeq& 1-6\epsilon+2\eta,
\label{ns}\\
r&\simeq& 16\epsilon.
\label{r}
\enann

Applying them to the present model, the power spectrum of the density  perturbations is written by
\begin{equation}
\mathcal{P}_\zeta\simeq \frac{\lambda \phi^6\left(4 M^2 +4\xi M^2 (6\xi-1)\phi+ \lambda \phi^4 \right)}{3072\pi^2  M^2(1-\xi\phi^2)^2}\,.
\label{cP}
\end{equation}
Since $U(\Phi)$ is not explicitly given by $\Phi$ but by some parametric representation of $\phi$, 
we have also used the original Higgs field $\phi$ 
as a ``parameter'' when we solve the dynamics.
\begin{widetext}
The e-folding number $N$ is given by
\bea
N&\simeq -{\displaystyle \int_{\Phi_{\rm end}}^{\Phi_N}{U\over dU/d\Phi}d\Phi}
&\simeq -\frac{\phi_N^2 \xi \{ 8 M^2 \xi^2(6\xi-1)+\lambda (2+\phi^2_N \xi )\}+2(\lambda +24M^2 \xi^3 )\ln \left[ 1-\xi {\phi^2_N}\right]}{64M^2\xi^3},
~~
\label{cN}
\ena
\end{widetext}
where $\Phi_N$ and $\phi_N$ are the values of $\Phi$ and $\phi$ 
at the e-folding number $N$, and $\Phi_{\rm end}$ and $\phi_{\rm end}$ are those values at the end of inflation ($\epsilon=1$).

\begin{figure}[h]
 \begin{center}
 \includegraphics[width=6cm]{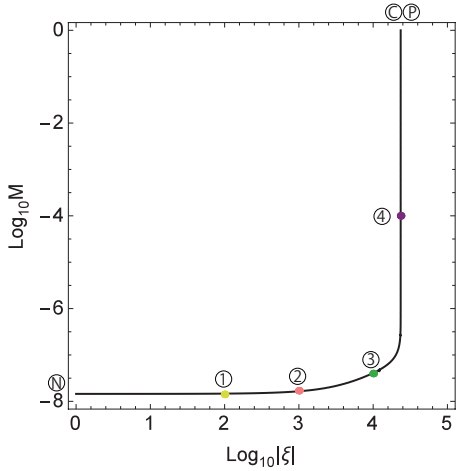}
\caption{The constraint on the coupling parameters  $M$ and $\xi$. 
The color dots with numbers or letters correspond to the coupling parameters given in 
Table. \ref{error}.  }
 \end{center}
  \label{fig:5}
\end{figure}

From the Planck CMB observation \cite{ade2016planck}, we have 
the constraint on the power spectrum
 $\mathcal{P}_\zeta\sim 2.2\times10^{-9}$.
Here we assume that $\Phi_{\rm end},\phi_{\rm end}$ are 
much smaller than $\Phi_N,\phi_N$, and then we set  $\Phi_{\rm end},\phi_{\rm end}=0$.
From Eq. (\ref{cN}), we obtain the value of the inflaton at $N=50-60$.
Substituting these values into Eq. (\ref{cP}), we obtain the constraint on $\xi$ and $M$, which is shown by the black curve in Fig. 5.

We then calculate $n_s$ and $r$, which results are plotted  
by the colored lines with numbers or letters in Fig. \ref{ns-r}.
We choose 7 sets of the coupling parameters:\\ 
~{\bf \textcircled{\scriptsize P}} (orange), 
{\bf \textcircled{\scriptsize N}} (blue), 
{\bf \textcircled{\scriptsize 1}} (yellow), 
{\bf \textcircled{\scriptsize 2}} (pink), \\
~{\bf \textcircled{\scriptsize 3}} (green), 
{\bf \textcircled{\scriptsize 4}} (purple), 
{\bf \textcircled{\scriptsize C}} (red),\\
which parameter values are given in Table \ref{error}.
The numbers or letters also correspond to the colored dots in Fig. 5. 
{\bf \textcircled{\scriptsize C}}, {\bf \textcircled{\scriptsize N}}, and {\bf \textcircled{\scriptsize P}}
give the models of 
the conventional Higgs inflation, of new Higgs inflation, and with a positive non-minimal coupling ($\xi=1/6$), respectively.

We find
that the predicted point ($n_s, r$) in the hybrid Higgs inflation  model
moves from  the values in new Higgs inflation  model to those in the conventional Higgs inflationary model as $|\xi|$ increases.
The tensor-to-scalar ratio $r$ changes drastically,  while the primordial tilt $n_s$ barely changes.
Hence once we know the tensor-scalar ratio $r$ by the future observations, 
we can fix the coupling parameters $M$ and $\xi$.

\begin{widetext}
\begin{center}
\begin{figure}[h]
 \includegraphics[width=10cm]{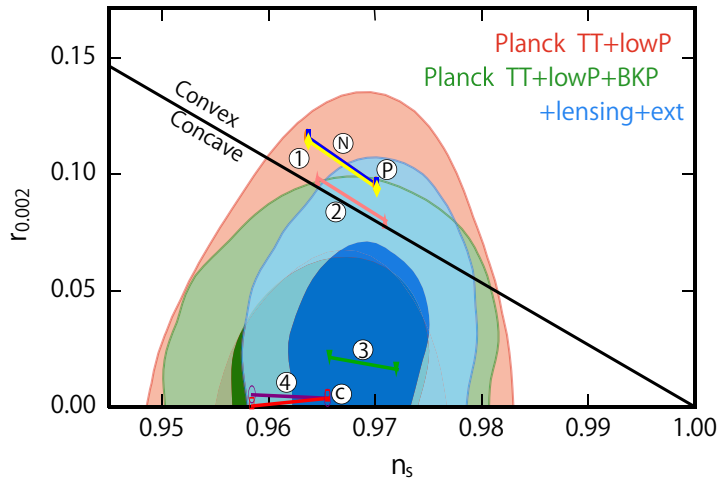}
  \caption{
 The colored lines show the numerical results of the hybrid Higgs inflation with $N=50-60$.
 The numbers or letters of lines correspond to the colored dots  in Fig. 5
 and the coupling parameters in Table  \ref{error}. 
\textcircled{\scriptsize N} and \textcircled{\scriptsize C} 
correspond to  new Higgs inflation model with  $M=1.41\times 10^{-8}$
 and the conventional Higgs inflation with  $ \xi=-10^{4}$, respectively. 
\textcircled{\scriptsize P} shows the case with positive non-minimal coupling
($\xi=1/6$), which line is overlapping with the line \textcircled{\scriptsize N} .
 The observational constraints on $n_s$ and $r$ by Planck 2015\cite{ade2016planck}
are also shown.}
 \label{ns-r}
\end{figure}
\end{center}

\end{widetext}

As found in Fig. \ref{ns-r}, when we have the derivative coupling, even if there exists 
a positive non-minimal coupling constant $(\xi>0)$, e.g., a conformal coupling ($\xi=1/6$), 
an inflationary scenario consistent with observations becomes possible\cite{Maeda:1988rb, futamase1989chaotic}.
The results are almost the same as new Higgs inflation model because non-minimal coupling 
does not play any important role except for the constraint on the initial value of the Higgs field
($\phi_{\rm in}<\xi^{-1/2}$).
The large positive coupling may not give a successful inflation.

\begin{widetext}

\begin{table}[!htb]
  \begin{center}
  \begin{tabular}{c|c|c||c|cc|cc}
  \hline
 &\multicolumn{2}{c||}{~}&&&&&\\[-0.7em]
&\multicolumn{2}{c||}{coupling parameters}&
\lower1ex\hbox{approximation }
&\multicolumn{2}{c|}{${\displaystyle{\Delta n_s/ n_s}}$}&\multicolumn{2}{c}{${\displaystyle{\Delta r / r}}$}\\
\cline{2-3}\cline{5-8}
&&&&&&&\\[-0.7em]
&$M$&$\xi$&\raise1ex\hbox{methods}&$N=60$&$N=50$&$N=60$&$N=50$\\
  \hline
&&&&&&&\\[-0.7em]
\textcircled{\scriptsize P}&$1.41\times 10^{-8}$&$1/6$&THD&$~7.1\times10^{-4}~$&$~1.0\times10^{-3}~$&$~3.5\times10^{-2}~$&$~4.5\times10^{-2}~$\\
                   &  &  &LFA&$~5.5\times10^{-4}~$&$~7.5\times10^{-4}~$&$~3.1\times10^{-2}~$&$~3.1\times10^{-2}~$\\
  \hline 
&&&&&&&\\[-0.7em]
\textcircled{\scriptsize N}&$1.41\times 10^{-8}$&$0$&THD&$~5.8\times10^{-2}~$&$~9.3\times10^{-4}~$&$~3.4\times10^{-2}~$&$~4.2\times10^{-2}~$\\
                   &  &  &LFL&$~5.0\times10^{-2}~$&$~7.6\times10^{-4}~$&$~3.1\times10^{-2}~$&$~3.8\times10^{-2}~$\\
  \hline 
 & &&&&&&\\[-0.7em]
 \textcircled{\scriptsize 1}& $1.41\times 10^{-8}$&$ -10^{2}$&THD&$7.5\times10^{-2}$&$1.1\times10^{-3}$&$3.6\times10^{-2}$&$4.5\times10^{-2}$\\
                &   &  &LFL&$6.0\times10^{-2}$&$7.9\times10^{-4}$&$3.1\times10^{-2}$&$3.8\times10^{-2}$\\
  \hline &&&&&&&\\[-0.7em]
\textcircled{\scriptsize 2}&  $1.70\times 10^{-8}$&$ -10^{3}$&THD&$7.6\times10^{-2}$&$1.1\times10^{-3}$&$3.6\times10^{-2}$&$4.6\times10^{-2}$\\
                   &  &  &LFL&$6.5\times10^{-4}$&$8.6\times10^{-4}$&$3.2\times10^{-2}$&$3.9\times10^{-2}$\\
  \hline &&&&&&&\\[-0.7em]
 \textcircled{\scriptsize 3}& $3.98\times 10^{-8}$&$ -10^{4}$&THD&$1.2\times10^{-3}$&$1.8\times10^{-3}$&$5.1\times10^{-2}$&$6.5\times10^{-2}$\\
                    &  & &LFL&$2.8\times10^{-3}$&$3.8\times10^{-3}$&$9.0\times10^{-2}$&$1.0\times10^{-1}$\\
  \hline &&&&&&&\\[-0.7em]
\textcircled{\scriptsize 4}&  $10^{-4}$&$ -4.27\times 10^{5}$&THD&$3.6\times10^{-3}$&$9.3\times10^{-4}$&$3.5\times10^{-2}$&$4.4\times10^{-2}$\\
&&&LFL&$4.1\times10^{3}$&$5.1\times10^{3}$&$1.1\times10^{5}$&$1.1\times10^{5}$\\
  \hline
  \end{tabular}
 \end{center}
\caption{The coupling parameters for which we evaluate the cosmological perturbations 
in the hybrid Higgs inflation model. We also show the
 accuracies of our results in our  truncated model 
in the Einstein frame (THD) and those in the large friction limit (LFL) in Jordan frame.
The accuracies in THD are always fairly good, while the accuracy in LFL becomes much worse 
as the coupling parameters approach the "conventional" Higgs inflation model, e.g. the model \textcircled{\scriptsize 4}, 
 because the approximation is no longer valid.}
 \label{error}
\end{table}
\end{widetext}

\subsection{Comparison with the ``exact'' results}
\label{Comparison}

In order to justify our results calculated in the truncated model, we have to evaluate the perturbations in the original full theory.
Since the basic equations in the Einstein frame is too complicated, we 
perform the calculation of perturbations in the original Jordan frame 
by use of the ADM formalism, which is summarized in Appendix \ref{ADM formalism}.
We suppose that this numerical result gives the correct values.
We then compare our results with those ``exact'' ones.
In Table \ref{error}, we show the ``accuracies'' of our results,
 which are evaluated by  the following deviations;
\beann
{\Delta n_s\over n_s}&:= &\frac{|n_s-n_{s,{\rm ADM}}|}{n_{s,{\rm ADM}}} \,,
\enann
\beann
{\Delta r\over r}&:= &\frac{|r-r_{\rm ADM}|}{r_{\rm ADM}}\,,
\enann
where 
$n_{s,{\rm ADM}}$ and $r_{\rm ADM}$ are the values obtained  in the full theory by use of the ADM formalism.

Our model is included in the class of generalized Higgs inflation models\cite{kamada2012generalized}.
In their model, they provide one simplified formula under some approximation.
Using it, we also calculate $n_s$ and $r$ in our model,
which are given in Appendix \ref{large_friction}. The accuracies of 
the perturbations are summarized in Table \ref{error}.
As the coupling parameters approach the region near the ``conventional" Higgs inflation, 
the deviations from the ``correct'' values increase.
The approximation used in \cite{kamada2012generalized} is broken near the ``conventional" Higgs inflation model.
On the other hand, our approximation method is still valid even in such parameter regions.

\section{Summary and Discussion}

In this paper we propose a hybrid type of the conventional Higgs inflation and new Higgs inflation models.
Performing a disformal transformation into the Einstein frame and truncating the higher-derivative terms
of the Higgs field,
we  analyze the inflationary background dynamics and the 
cosmological perturbations.
From the observed power spectrum of the density perturbations, 
we show the constraint on  the non-minimal coupling constant $\xi$ and the mass parameter $M$
 in the derivative coupling. 
Although the primordial tilt $n_s$ in the hybrid model barely changes,  the tensor-to-scalar ratio $r$ 
moves from  the value in new Higgs inflationary model to that in the conventional Higgs inflationary model as 
 $|\xi|$ increases. $r$ varies in the range of the observationally arrowed region.
 Hence once we know the tensor-scalar ratio $r$ by the future observations, 
we can fix the coupling parameters $M$ and $\xi$.
We confirm our results by numerical analysis by ADM formalism of the full theory in the Jordan frame.
The higher-derivative terms of the Higgs field appeared in the Einstein fame can become negligible 
because of  the slow-rolling evolution during inflation.

As we mentioned in Introduction, our studied model is no more than the tree level.
At high energy scale such as an inflationary stage, we should take into account the quantum loop effects 
to obtain the effective potential.
Since the loop corrections change the effective Higgs potential via the running of the coupling constants very much, 
the conventional Higgs inflation model has been analyzed
 including in the loop corrections\cite{hamada2015higgs,bezrukov2015living,DeSimone:2008ei}.
The frame dependence of the effective potential has also been discussed\cite{hamada2015higgs,Bezrukov:2009db}.
This problem is caused by  in which frame (either in the Jordan frame or in the Einstein frame) 
we quantize the Higgs field and calculate the effective potential.
They found that  $r$ would be enhanced so much that some parameter regions can be
 excluded by the observations.
 However, as we expect, the result should not depend on the frame choice\cite{George:2013iia,Hamada:2016onh} .
Since we also have the stability problem of the Higgs field, 
we have to consider the loop corrections more carefully.
We expect not only the Higgs coupling $\lambda$  but also the non-minimal coupling $\xi$
 as well as the derivative coupling $M$ are running via renormalization.
Such a calculation could be performed by the asymptotic safety approach
\cite{Xianyu2014asymptotic, Oda2016asymptotic, Hamada:2017rvn,Atkins:2010yg}.
These problems are under investigation.


\section*{Acknowledgments}
KM would like to thank Department of Physics, University of Auckland for their hospitality during his visit in 2016, where the preliminary work was started.
He also acknowledge Richard Easther and Nathan Musoke for  the useful discussions.
SS  would like to thank  Hiroyuki Abe, Katsuki Aoki,  Mitsuhiro Fukushima, Yuta Hamada, Kohei Kamada and Yosuke Misonoh for valuable comments and  fruitful discussions.
This work was supported in part by  JSPS KAKENHI Grant Numbers
 JP16K05362  and JP17H06359. 



\clearpage

\appendix

\section{Towards the Einstein Frame via Disformal Transformation}
\label{Towards_Einstein}
\subsection{Disformal transformation}
\label{disformal_transformation}

We first consider the following  disformal transformation 
\bea
g_{\mu\nu}=\bar \Omega^2\left(\bar g_{\mu\nu}+\bar u_\mu \bar u_\nu \right)
\label{disformal_transformation}
\ena
 from the 
original metric $\bar g_{\mu\nu}$ with a  vector field $ \bar u^\mu$, which 
 is either timelike or spacelike, i.e.,
\beann
\bar u^\mu \bar u_\mu=\epsilon \bar u^2
\,,
\enann
where $\epsilon=\pm 1$.
This transformation leads
\beann
\sqrt{- g}&=&\bar \Omega^D(1+\epsilon \bar u^2)^{1\over 2}\sqrt{-\bar g}
\nn
g^{\mu\nu}&=&\bar \Omega^{-2}\left(\bar g^{\mu\nu}-{1\over 1+\epsilon\bar u^2}\bar u^\mu \bar u^\nu
\right)
\nn
\Gamma^\mu_{\nu\rho}&=&\bar \Gamma^\mu_{\nu\rho}+\bar \gamma^\mu_{\nu\rho}
\,,
\enann
where $D$ is the spacetime dimension, and 
the deviation of the connection $\bar \gamma^\mu_{\nu\rho}$ is given by 
\bea
\bar \gamma^\mu_{\nu\rho}=\bar f^\mu_{\nu\rho}+\bar \omega^\mu_{\nu\rho}
\ena
with 
\begin{widetext}
\beann
\bar f^\mu_{~\rho\sigma}&:=&
{1\over 2}\left(\bar g^{\mu\nu}-{1\over 1+\epsilon\bar u^2}\bar u^\mu \bar u^\nu\right)
\left[\bar \nabla_\rho(\bar u_\nu \bar u_\sigma)+\bar \nabla_\sigma(\bar u_\nu \bar u_\rho)
-\bar \nabla_\nu(\bar u_\rho \bar u_\sigma)
\right]
\nn
\bar \omega^\mu_{~\rho\sigma}&:=&
\delta^\mu_{\rho}\bar \nabla_\sigma \ln \bar \Omega
+\delta^\mu_{\sigma}\bar \nabla_\rho \ln \bar \Omega
-\left(\bar g^{\mu\nu}-{1\over 1+\epsilon\bar u^2}\bar u^\mu \bar u^\nu\right)
(\bar g_{\rho\sigma}+\bar u_\rho \bar u_\sigma)\bar \nabla_\nu \ln \bar \Omega
\,,
\enann
where $\bar\nabla^\mu$ denotes the covariant derivative with respect to the metric $\bar g_{\mu\nu}$.

The Riemann, Ricci and scalar curvatures are given by 
\beann
R^\mu_{~\nu\rho\sigma}&=&\bar R^\mu_{~\nu\rho\sigma}
+\bar \nabla_\rho \bar \gamma^\mu_{~\nu\sigma}
-\bar \nabla_\sigma \bar \gamma^\mu_{~\nu\rho}
+\bar \gamma^\mu_{~\alpha\sigma} \bar \gamma^\alpha_{~\nu\rho}
-\bar \gamma^\mu_{~\alpha\rho} \bar \gamma^\alpha_{~\nu\sigma}
\nn
R_{\rho\sigma}&=&\bar R_{\rho\sigma}
+\bar \nabla_\mu \bar \gamma^\mu_{\rho\sigma}
-\bar \nabla_\sigma \bar \gamma^\mu_{~\mu\rho}
+\bar \gamma^\mu_{~\mu\alpha} \bar \gamma^\alpha_{\rho\sigma}
-\bar \gamma^\beta_{~\alpha\rho} \bar \gamma^\alpha_{~\beta\sigma}
\nn
R
&=&\bar \Omega^{-2}\Big[{2+\epsilon\bar u^2\over 2(1+\epsilon\bar u^2)}\bar R
-{1\over  1+\epsilon\bar u^2}\bar G_{\mu\nu}\bar u^\mu \bar u^\nu
+\bar \nabla_\mu (\bar g^{\rho\sigma} \bar \gamma^\mu_{\rho\sigma})
-\bar \nabla^\rho \bar \gamma^\mu_{~\mu\rho}
+\bar g^{\rho\sigma}\bar \gamma^\alpha_{\rho\sigma}\bar \gamma^\mu_{~\mu\alpha} 
-\bar g^{\rho\sigma}\bar \gamma^\beta_{~\alpha\rho} \bar \gamma^\alpha_{~\beta\sigma}
\nn
&& 
-{1\over 1+\epsilon\bar u^2}\bar u^\rho \bar u^\sigma
\Big(\bar \nabla_\mu \bar \gamma^\mu_{\rho\sigma}
-\bar \nabla_\sigma \bar \gamma^\mu_{~\mu\rho}
+\bar \gamma^\mu_{~\mu\alpha} \bar \gamma^\alpha_{\rho\sigma}
-\bar \gamma^\beta_{~\alpha\rho} \bar \gamma^\alpha_{~\beta\sigma}
\Big)\Big]
\,.
\enann
Hence we obtain the Einstein tensor as
\beann
G_{\mu\nu}
&=&
\bar G_{\mu\nu}+{1\over 2(1+\epsilon\bar u^2)}
(\bar g_{\mu\nu}+\bar u_\mu \bar u_\nu)\bar u^\alpha \bar u^\beta \bar G_{\alpha\beta}
+{\epsilon\bar u^2\over 4(1+\epsilon\bar u^2)}(\bar g_{\mu\nu}+\bar u_\mu \bar u_\nu)\bar R
-{1\over 2}\bar u_\mu \bar u_\nu \bar R
\nonumber \\
&&+\bar \nabla_\rho \bar \gamma^\rho_{\mu\nu}
-\bar \nabla_\nu \bar \gamma^\rho_{~\rho\mu}
+\bar \gamma^\rho_{~\rho\sigma} \bar \gamma^\sigma_{\mu\nu}
-\bar \gamma^\sigma_{~\rho\mu} \bar \gamma^\rho_{~\sigma\nu}
\nn
&&-{1\over 2}(\bar g_{\mu\nu}+\bar u_\mu \bar u_\nu)
\Big[\bar \nabla_\alpha (\bar g^{\rho\sigma} \bar \gamma^\alpha_{\rho\sigma})
-\bar \nabla^\rho \bar \gamma^\sigma_{~\sigma\rho}
+\bar g^{\rho\sigma}\bar \gamma^\alpha_{\rho\sigma}\bar \gamma^\beta_{~\beta\alpha} 
-\bar g^{\rho\sigma}\bar \gamma^\beta_{~\alpha\rho} \bar \gamma^\alpha_{~\beta\sigma}
\nn
&& 
-{1\over 1+\epsilon\bar u^2}\bar u^\rho \bar u^\sigma
\Big(\bar \nabla_\alpha \bar \gamma^\alpha_{\rho\sigma}
-\bar \nabla_\sigma \bar \gamma^\alpha_{~\alpha\rho}
+\bar \gamma^\beta_{~\beta\alpha} \bar \gamma^\alpha_{\rho\sigma}
-\bar \gamma^\beta_{~\alpha\rho} \bar \gamma^\alpha_{~\beta\sigma}
\Big)\Big]
\enann

\subsection{Towards the Einstein frame}
\label{Einstein_frame}
Now we consider the following action
\bea
S=\int d^D x \sqrt{-g} \Big[F_4(\phi) R(\tilde g) +F_5(\phi)
G^{\mu\nu}(g)\nabla_\mu\phi \nabla_\nu\phi\Big]
\label{actionF4F5}
\,,
\ena
where 
$F_4(\phi) $ and $F_5(\phi) $ are arbitrary  functions of a scalar field $\phi$.
This gives a gravity sector of some simple 
 scalar-tensor gravity theory.

Performing the disformal transformation 
(\ref{disformal_transformation})
with 
$
\bar u_\mu=\beta \bar \nabla_\mu \phi
\,,
$
we find
\beann
R
&=&\bar \Omega^{-2}\Big[{2+\epsilon\bar u^2\over 2(1+\epsilon\bar u^2)}\bar R
-{1\over  1+\epsilon\bar u^2}\bar G_{\mu\nu}\bar u^\mu \bar u^\nu
+\cdots \Big]
\enann

\beann
G_{\mu\nu}
&=&
\bar G_{\mu\nu}+{1\over 2(1+\epsilon\bar u^2)}
(\bar g_{\mu\nu}+\bar u_\mu \bar u_\nu)\bar u^\alpha \bar u^\beta \bar G_{\alpha\beta}
-{\bar u^2\over 4(1+\epsilon\bar u^2)}(\bar g_{\mu\nu}+\bar u_\mu \bar u_\nu)\bar R
-{1\over 2}\bar u_\mu \bar u_\nu \bar R
+\cdots
\,,
\enann
where $\cdots$ describes some functions of $\phi$ and its derivatives.

Since $\nabla_\mu\phi=\partial_\mu\phi= \bar \nabla_\mu\phi$,
we find the action (\ref{actionF4F5}) as
\beann
S
&=&
\int d^D x \sqrt{-\bar g}\Big[A (\bar \Omega, \beta) \bar R(\bar g)+ B(\bar \Omega, \beta) \bar G_{\alpha\beta}(\bar g) \bar \nabla^\alpha \phi \bar \nabla^\beta \phi + \cdots\Big]
\,,
\enann
where
\beann
A&=&{\bar \Omega^{D-4}(1+\epsilon\bar u^2)^{-3/2}\over 4\beta^2}\left[
2\beta^2\bar \Omega^2(1+\epsilon\bar u^2)(2+\epsilon\bar u^2)F_4-(\epsilon\bar u^2)^2 F_5\right]
\\
B&=&{\bar \Omega^{D-4}(1+\epsilon\bar u^2)^{-3/2}\over 2}\left[-
2\beta^2\bar \Omega^2(1+\epsilon\bar u^2)F_4+(2+\epsilon\bar u^2) F_5\right]
\,.
\enann

In order to find only the Einstein-Hilbert action in the $\bar g$ frame,
we have to impose 
$A={1 \over 2}$ and $B=0$.
We then find the equations for $\bar \Omega$ and $\beta$ as
\beann
&&
2\beta^2\bar \Omega^2(1+\epsilon\bar u^2)(2+\epsilon\bar u^2)F_4-(\epsilon\bar u^2)^2 F_5
= 2
\beta^2\bar \Omega^{4-D}(1+\epsilon\bar u^2)^{3/2}
\\
&&-2\beta^2\bar \Omega^2(1+\epsilon\bar u^2)F_4+(2+\epsilon\bar u^2) F_5=0
\enann
\end{widetext}
These equations fix $\bar\Omega$ and $\beta$ as 
\bea
\bar \Omega^{D-2}&=&
{(2+\epsilon \bar u^2)\over 4 (1+\epsilon\bar u^2)^{1/2}F_4}
\label{Omega}
\\
\beta^2&=&{(2+\epsilon\bar  u^2)F_5\over 2(1+\epsilon\bar  u^2)F_4}\, \bar \Omega^{-2}
\,.
\label{beta}
\ena
Since $\epsilon \bar u^2=\beta^2(\nabla\phi)^2$, we have to solve the coupled 
equations (\ref{Omega}) and (\ref{beta}) 
to describe $\bar \Omega$ and $\beta$ explicitly in terms of $\phi$ and $\nabla\phi$.

As a result, the disformal transformation (\ref{disformal_transformation}) with (\ref{Omega}) and (\ref{beta}) gives the Einstein gravity in the $\bar g$ frame, which we call 
the Einstein frame.

\subsection{Hybrid Higgs Inflation}
\label{Hybrid Higgs Inflation}
Next we apply the above result to the hybrid Higgs inflation model in four dimensions
($D=4$), which action is given by
\beann 
S&=&
\int d^4 x \sqrt{-g}\Big[
{1-\xi\phi^2\over 2}R
\\
&&
-{1\over 2}
\left(g^{\mu\nu}-{G^{\mu\nu}\over M^2} \right)\nabla_\mu \phi\nabla_\nu \phi 
-V(\phi)
\Big]
\,.
\enann

Since we are interested in a cosmological dynamics, 
for the disformal transformation 
\beann
g_{\mu\nu}=\bar \Omega^2\left(\bar g_{\mu\nu}+\beta^2 \bar \nabla_\mu \phi \bar \nabla_\nu\phi\right)
\,,
\enann
we 
choose a timelike vector $\bar u_\alpha=\beta \bar\nabla_\mu \phi$
 with $\bar u^\alpha \bar u_\alpha=-\bar u^2$($\epsilon=-1$).
 
To find the Einstein gravity in the transformed $\bar g$ frame, 
we choose 
\beann
\bar \Omega^{2}
&=& {(2-\bar u^2)\over 2(1-\bar u^2)^{1\over 2}(1-\xi\phi^2)}
\\
\beta^2&=&
{1\over 
(1-\bar u^2)^{1\over 2} M^2}\,.
\enann

Introducing the canonical kinetic term of the Higgs field
\beann
\bar X:=-{1\over 2}\bar g^{\mu\nu}\bar \nabla_\mu \phi \bar \nabla_\nu \phi 
\,,
\enann
since $\bar u^2=-\bar u^\alpha \bar u_\alpha=2\beta^2 \bar X$, 
we find the relation between $\bar u$ and $\bar X$ as
\bea
\bar u^2(1-\bar u^2)^{1\over 2}
={2\bar X\over M^2 }
\label{lambda_X}
\,.
\ena

The gravity action is now given by the Einstein-Hilbert term.
How about the action of  the Higgs field?
Apart from the standard action in the original frame, 
we have the additional contribution $L_{\rm disformal}$
 from the disformal transformation
 as
 \begin{widetext}
 \beann
S_{\rm Higgs}&=&\int d^4x\sqrt{-g}\Big[
-{1\over 2}
g^{\mu\nu}\nabla_\mu \phi \nabla_\nu \phi 
-V(\phi)
+L_{\rm disformal}
\Big]\,,
\enann
where 
\beann
L_{\rm disformal}&=&{1-\xi\phi^2\over 2\bar \Omega^{2}}\Big[
\bar \nabla_\mu (\bar g^{\rho\sigma}\bar  \gamma^\mu_{\rho\sigma})
-\bar \nabla^\rho \bar \gamma^\mu_{~\mu\rho}
+\bar g^{\rho\sigma}\bar \gamma^\alpha_{\rho\sigma}\bar \gamma^\mu_{~\mu\alpha} 
-\bar g^{\rho\sigma}\bar \gamma^\beta_{~\alpha\rho} \bar \gamma^\alpha_{~\beta\sigma}
\nn
&& 
-{\beta^2\over 1-\bar u^2}\bar \nabla^\rho \phi \bar \nabla^\sigma\phi
\Big(\bar \nabla_\mu \bar \gamma^\mu_{\rho\sigma}
-\bar \nabla_\sigma \bar \gamma^\mu_{~\mu\rho}
+\bar \gamma^\mu_{~\mu\alpha} \bar \gamma^\alpha_{\rho\sigma}
-\bar \gamma^\beta_{~\alpha\rho} \bar \gamma^\alpha_{~\beta\sigma}
\Big)\Big]
\\
&&
+{\bar g^{\mu\lambda}\bar g^{\nu\kappa}\bar \nabla_\lambda \phi  \bar \nabla_\kappa \phi \over 2M^2}
\Big[
\bar \nabla_\rho \bar \gamma^\rho_{\mu\nu}
-\bar \nabla_\nu \bar \gamma^\rho_{~\rho\mu}
+\bar \gamma^\rho_{~\rho\sigma} \bar \gamma^\sigma_{\mu\nu}
-\bar \gamma^\sigma_{~\rho\mu} \bar \gamma^\rho_{~\sigma\nu}
\nn
&&-{1\over 2}(\bar g_{\mu\nu}+\beta^2 \bar \nabla_\mu \phi \bar \nabla_\nu\phi)
\Big[\bar \nabla_\alpha (\bar g^{\rho\sigma} \bar \gamma^\alpha_{\rho\sigma})
-\bar \nabla^\rho \bar \gamma^\sigma_{~\sigma\rho}
+\bar g^{\rho\sigma}\bar \gamma^\alpha_{\rho\sigma}\bar \gamma^\beta_{~\beta\alpha} 
-\bar g^{\rho\sigma}\bar \gamma^\beta_{~\alpha\rho} \bar \gamma^\alpha_{~\beta\sigma}
\nn
&& 
-{\beta^2 \over 1-\bar u^2}\bar \nabla^\rho \phi \bar \nabla^\sigma\phi 
\Big(\bar \nabla_\alpha \bar \gamma^\alpha_{\rho\sigma}
-\bar \nabla_\sigma \bar \gamma^\alpha_{~\alpha\rho}
+\bar \gamma^\beta_{~\beta\alpha} \bar \gamma^\alpha_{\rho\sigma}
-\bar \gamma^\beta_{~\alpha\rho} \bar \gamma^\alpha_{~\beta\sigma}
\Big)
\Big]\,.
\enann

In order to obtain the action in the Einstein frame, 
we have to rewrite the variables in the $g$-frame to those in the $\bar g$-frame.
Since $L_{\rm disformal}$ is already given by the variables in the  $\bar g$-frame,
we easily find the action of the Higgs field in the Einstein frame as
\beann
S_{\rm Higgs}&=&\int d^4x\sqrt{-\bar g}\bar \Omega^4(1-\bar u^2)^{1\over 2}
\Big[
-{1\over 2}
\bar\Omega^{-2}\bar g^{\mu\nu}\bar \nabla_\mu \phi 
\bar \nabla_\nu \phi 
-V(\phi)
+L_{\rm disformal}
\Big]\,.
\enann

\end{widetext}
Since the above additional term $L_{\rm disformal}$ is very complicated, 
it seems not to be useful for a practical purpose.
However, since we are interested in an inflationary era, at which the scalar field 
changes very slowly, the kinetic term $\bar X$ is small in the slow-rolling phase
and then the higher-derivative terms such as ${\displaystyle \bar {\Box}} \phi$ and $\bar X^2$ may be able to ignored.
The resultant action becomes much simpler.
We shall show  below what we find as the lowest order.

Expanding Eq. (\ref{lambda_X}) in terms of $\bar X$,
we find 
\beann
\bar u^2={2\bar X\over M^2 }
\left(1+{\bar X\over M^2 } +\cdots\right)
\,.
\enann

The kinetic term and the potential term in the standard Higgs action are 
rewritten by the similar expansion as
\beann
&&
-{1\over 2}\sqrt{-g}(\nabla \phi)^2
\\
&&~= 
-{1\over 2}\sqrt{-g}(\bar \nabla \phi)^2\times 
{\left[1-{ (\bar \nabla \phi)^2 \over 2M^2}+\cdots \right]
\over 1-\xi\phi^2}
\\
\enann
\beann
&&
-\sqrt{-g}V(\phi)
\\
&&~= 
 -\sqrt{- \bar g}{V(\phi) \over (1-\xi\phi^2)^2}
 \left[
 1
 +
 { (\bar \nabla \phi)^2  \over 2M^2}-{(\bar \nabla\phi)^4\over 8M^4}\cdots
 \right]
\enann

The additional term $L_{\rm disformal}$, in which  both $\bar f^\mu_{~\nu\rho}$ and $\bar \omega^\mu_{~\nu\rho}$ 
are the functionals of $\phi$ and its derivatives, is very complicated.
However $\bar f^\mu_{~\nu\rho}$ consists only of the derivatives of $\bar u^\mu=\beta\bar \nabla^\mu \phi$, which are the higher-derivative terms of $\phi$. We may neglect such terms.
On th other hand,  $\bar \omega^\mu_{~\nu\rho}$ contains the derivative of the conformal factor 
$\bar \Omega$.
 Hence its contribution provides the square of the first derivative term of the scalar field $\phi$,
 which gives the lowest order in the expansion.  Hence we have to pick up the relevant contributions 
 from  the therms with $\bar \omega^\mu_{~\nu\rho}$. 
 Since the derivatives of $\bar \omega^\mu_{~\nu\rho}$
  are the higher derivatives of $\phi$, only algebraic terms remain.

We then find
\beann
&&
\sqrt{-g}L_{\rm disformal}
\\
&&~=\sqrt{- \bar g}\bar \Omega^4(1- \bar u^2)^{1\over 2}{1-\xi\phi^2\over 2\bar \Omega^{2}}
\\
&&~~\times
\Big[\bar g^{\rho\sigma}\bar \omega^\alpha_{~\rho\sigma}\bar \omega^\mu_{~\mu\alpha}
-\bar g^{\rho\sigma}\bar \omega^\beta_{~\alpha\rho}\bar \omega^\alpha_{~\beta\sigma}+\cdots\Big]
\\
&&~=\sqrt{- \bar g}\left[-3{\xi^2 \phi^2\over ( 1-\xi\phi^2)^2}(\bar \nabla\phi)^2 +\cdots\right]
\enann

As a result, the effective lowest-order action for the Higgs field in the Einstein frame is given by
\begin{widetext}
\beann
S_{\rm Higgs}&=&-\int d^4x
\sqrt{-\bar g}
\left[{1\over 2}\left(
{1-\xi(1-6\xi)\phi^2+  {V(\phi) \over M^2 }\over 
 \left(1-\xi\phi^2\right)^2}
 \right)(\bar \nabla \phi)^2
+
{V(\phi) \over \left(1-\xi\phi^2\right)^2}
+\cdots  \right]
\nn
&\approx&
-\int d^4x
\sqrt{-\bar g}\left[{1\over 2}(\bar \nabla \Phi)^2
+
U(\Phi) 
+\cdots  \right]
\,,
\enann
where
\beann
{d\Phi\over d\phi}&=&{\sqrt{1-\xi(1-6\xi)\phi^2+  {V(\phi) \over M^2 }}\over 1-\xi\phi^2}
\nn
U(\Phi)&=&{V(\phi) \over  \left(1-\xi\phi^2\right)^2}
\enann

\section{Analysis in Jordan Frame}
\label{Jordan_frame}

\subsection{The basic equations for Friedmann universe in Jordan frame}
\label{basic_equations_Jordan}
We shall derive the basic equations for a background universe in Jordan frame.
We adopt the flat FLRW metric as 
\beann
ds^2=- {\cal N}^2dt^2+a(t)^2d\vect{x}^2\,,
\label{flrw}
\enann
where ${\cal N}$ is a lapse function and $a$ is a scale factor of the universe.
The action (\ref{hha}) is written as
\beann
S=\int d^4x {\cal N}a^3
\left[ -3{H^2\over {\cal N}^2}\left(1-\xi \phi^2\right) +\left( 1+\frac{3H^2}{{\cal N}^2M^2}\right) \frac{\dot{\phi}^2}{2{\cal N}^2}+6\xi {H\phi\dot{\phi}
\over {\cal N}^2}-V(\phi) \right],
\enann
where $H:=\dot{a}/a$ is the Hubble expansion  parameter and a dot denotes 
the derivative with respect to the cosmic time $t$.
\par
Taking the variation with respect to the lapse function ${\cal N}$, we find 
\begin{equation}
H^2=\frac{1}{3(1-\xi \phi^2)}\left[\left(1+\frac{9H^2}{M^2}\right)\frac{\dot{\phi}^2}{2}+6\xi \phi \dot{\phi}H+V(\phi)\right],
\label{fr1}
\end{equation}
where we have set ${\cal N}=1$.
This is just the Friedmann equation in the Jordan frame. 
The variation with respect to the scale factor $a$ gives 
\begin{equation}
\left( 3H^2+2\dot{H}\right)\left(1-\xi\phi^2 \right) +\left(1-\frac{3H^2}{M^2}-4\xi\right)\frac{\dot{\phi}^2}{2}-V(\phi)-\frac{1}{M^2}\frac{d}{dt}(H\dot{\phi}^2)-2\xi\phi\ddot{\phi}-4\xi\phi\dot{\phi}H=0
.
\label{fr2}
\end{equation}
The equation of motion of the Higgs field is given by variation with respect to $\phi$ as
\begin{equation}
\left( 1+\frac{3H^3}{M^2}\right)\ddot{\phi}+3H\left( 1+\frac{3H^2+2\dot{H}}{M^2}\right)\dot{\phi}+6\xi \left( \dot{H}+2H^2\right)\phi+ \frac{dV}{d\phi}=0.
\label{eom}
\end{equation}
Eqs. (\ref{fr1}), (\ref{fr2}) and (\ref{eom}) are the basic equations for Friedmann universe in the Jordan frame.

\subsection{Cosmological Perturbations in Jordan frame by use of the ADM formalism}
\label{ADM formalism}
In order to check the accuracy of the perturbations calculated 
in the truncated model in the Einstein frame, 
we have to analyze the perturbations in the full theory.
Although the disformal transformation provides the equivalent cosmological perturbations\cite{makino1991density, minamitsuji2014disformal, tsujikawa2015disformal, watanabe2015multi, domenech2015cosmological, motohashi2016disformal},
 the full action in the Einstein frame is too complicated to be analyzed. 
Hence we perform the perturbations in the original Jordan frame by use of the ADM formalism\cite{kobayashi2011generalized} .
 We expect that this approach gives the correct perturbations.
We then compare those with our results evaluated by the truncated model in the Einstein frame.

We use the unitary gauge, in which Higgs field is uniform $\phi=\phi(t)$.
The perturbed metric is written as
\beann
ds^2 =-{\cal N}^2 dt^2+\gamma_{ij}\left(dx^i+{\cal N}^i dt\right)\left(dx^j+{\cal N}^j dt\right) 
\,,
\enann
where
\beann
{\cal N}=1+\alpha ,~~~~{\cal N}_i=\partial_i \beta ,~~~~\gamma_{ij}=a^2(t)e^{2\zeta}\left(\delta_{ij}+h_{ij}+\frac{1}{2}h_{ik}h_{kj}\right).
\enann
$\alpha,\beta,\zeta$ and $h_{ij}$ are the scalar modes and 
 the tensor mode of perturbations, respectively. $h_{ij}$ satisfies the transverse-traceless conditions $h_{ii}=h_{ij,j}=0$, 

First we consider the scalar perturbations.
Varying the action with respect to the lapse $\alpha$ and the shift $\beta$,
we find two constraint equations,
\beann
\alpha&=&\frac{1-\xi\phi^2-\frac{\dot{\phi}^2}{2M^2}}{H\left(1-\xi{\phi^2}-\frac{3\dot{\phi}^2}{2M^2}\right)-\xi\phi\dot{\phi}}\dot{\zeta},\\
\frac{\partial_i^2 \beta}{a^2}&=&-\frac{1-\xi\phi^2-\frac{\dot{\phi}^2}{2M^2}}{H\Big(1-\xi\phi^2-\frac{3\dot{\phi}^2}{2M^2}\Big)-\xi\phi\dot{\phi}}\frac{\partial^2_i\zeta}{a^2}\\
&&+\left\{3+\frac{1-\xi{\phi^2}-\frac{\dot{\phi}^2}{2M^2}}{\left[ H
\left(1-\xi{\phi^2}-\frac{3\dot{\phi}^2}{2M^2}\right)-\xi{\phi\dot{\phi}}\right]^2}\left[\frac{\dot{\phi}^2}{2}+{6H\xi \phi\dot{\phi}}-3H^2\left( 1-\xi{\phi^2}-\frac{3\dot{\phi}^2}{M^2}\right)
\right]
\right\}\dot{\zeta}
\,.
\enann
Replacing  $\alpha$ and $\beta$ with $\zeta$ by these constraints, we obtain 
the quadratic action for the scalar perturbations as
\beann
S_S^{(2)}=\int dtd^3\vect{x}  ~a^3\left[ G_S\dot{\zeta}^2-\frac{F_S}{a^2}\left(\partial_i\zeta\right)^2\right]
\label{adm2ndS}
\enann
where,
\beann
G_S&=& 3\left(1-\xi{\phi^2}-\frac{\dot{\phi}^2}{2M^2}\right)
\nonumber
\\
&&+\frac{\left(1-\xi{\phi^2}-\frac{\dot{\phi}^2}{2M^2}\right)^2}{\left[ H\left(1-\xi{\phi^2}-\frac{3\dot{\phi}^2}{2M^2}\right)-\xi{\phi\dot{\phi}}\right]^2}\left[-3H^2\left( 1-\xi{\phi^2}-\frac{3\dot{\phi}^2}{M^2}\right)+\frac{\dot{\phi}^2}{2}+{6H\xi \phi\dot{\phi}}\right]\\
F_S&=&\frac{1}{a}\frac{d}{dt}\left[\frac{a\left(1-\xi{\phi^2}-\frac{\dot{\phi}^2}{2M^2}\right)^2}{H\left(1-\xi{\phi^2}-\frac{3\dot{\phi}^2}{2M^2}\right)
-\xi{\phi\dot{\phi}}}\right]-\left(1-\xi{\phi^2}+\frac{\dot{\phi}^2}{2M^2}\right)
\,.
\enann

\end{widetext}

The sound speed $c_S$  is given by
\beann
c_S^2=\frac{F_S}{G_S}.
\enann
We assume the following conditions during inflation:
\beann
\epsilon & :=& -\frac{\dot{H}}{H^2}\simeq {\rm const}\\
f_S & :=& \frac{\dot{F}_S}{HF_S}\simeq  {\rm const}\\
g_S & :=& \frac{\dot{G}_S}{HG_S}\simeq {\rm const}
\,,
\enann
which are confirmed by numerical analysis.
The power spectrum  of the scalar perturbations is given by
\bea
\mathcal{P}_\zeta=\left.\frac{\gamma_S}{2}\sqrt{\frac{G_S}{F_S^3}}\frac{H^2}{4\pi^2} \right|_{\rm horizon}\,,
\label{scalar_Jordan}
\ena
where
\begin{widetext}
\beann
\gamma_S=2^{\frac{2(3-\epsilon+g_S)}{2-2\epsilon-f_S/2+g_S/2}-3}\left|\frac{\Gamma\left(\frac{3-\epsilon+g_s}{2-2\epsilon-f_S/2+g_S/2}\right)}{\Gamma\left(\frac{3}{2}\right)}\right|^2 \left( 1-\epsilon-\frac{f_S+g_S}{2}\right)^2
\enann
\end{widetext}
$\Gamma$ is the Gamma function. 
The subscript ``horizon'' means that 
the  perturbations are evaluated at the crossing of  the sound horizon. 
The spectrum index is obtained by
\beann
n_s-1=3-\frac{2(3-\epsilon+g_s)}{2-2\epsilon-f_S/2+g_S/2}.
\label{Ans}
\enann
\par
The tensor perturbations are also calculated in the same manner as the scalar perturbations.
The quadratic action for the tensor perturbations is
\beann
S_T^{(2)}=\frac{1}{8}\int dt d^3\vect{x}~ a^3\left[ G_T\left( \dot{h}_{ij}\right)^2-\frac{F_T}{a^2}\left(\partial _k h_{ij}\right)^2\right]
\,,
\enann
where
\beann
G_T&=& 1-\xi {\phi^2}-\frac{\dot{\phi}^2}{2M^2}\\
F_T&=& 1-\xi {\phi^2}+\frac{\dot{\phi}^2}{2M^2}.
\enann
We also assume the following conditions:
\beann
f_T & =& \frac{\dot{F}_T}{HF_T}\simeq {\rm const}\\
g_T & =& \frac{\dot{G}_T}{HG_T}\simeq  {\rm const}
\,,
\enann
which are also confirmed numerically.

The power spectrum for tensor perturbations is given by
\bea
\mathcal{P}_T=8\gamma_T \left.\sqrt{\frac{G_T}{F_T^3}}\frac{H^2}{4\pi^2}\right|_{\rm horizon}
\label{tensor_Jordan}
\,,
\ena
where
\begin{widetext}
\beann
\gamma_T := 2^{\frac{2(3-\epsilon+g_T)}{2-2\epsilon-f_T/2+g_T/2}-3}\left|\frac{\Gamma\left(\frac{3-\epsilon+g_T}{2-2\epsilon-f_T/2+g_T/2}\right)}{\Gamma\left(\frac{3}{2}\right)}\right|^2 \left( 1-\epsilon-\frac{f_T+g_T}{2}\right)^2.
\enann
\end{widetext}

As a result, we obtain the tensor-to-scalar ratio $r$  by
\bea
r=\frac{\mathcal{P}_T}{\mathcal{P}_\zeta}
\,.
\label{Ar}
\ena

We expect that those values give the correct ones, although we have to perform numerical calculation.
We compare the results obtained by some approximations with the above ``correct'' values, and then judge 
the validity of the approximations.

\subsection{Large Friction Limit}
\label{large_friction}
Eqs. (\ref{scalar_Jordan}) and (\ref{tensor_Jordan}) are still complicated to be evaluated.
Hence, in \cite{kamada2012generalized}, taking the large friction limit, 
they have given a simplified formula.
Since our model is one of the generalized Higgs inflation models, 
we may apply their formula to our analysis.


Here we show the results for the hybrid Higgs inflation model by use of their formula.
The power spectra of the scalar and tensor perturbations  are given by 
\beann
\mathcal{P}_{\zeta} \simeq \frac{\lambda \phi^6}{768\pi^2 (1-\xi \phi^2)}\left(1+\frac{\lambda \phi^4}{4 M^2(1-\xi \phi^2)}\right)\,,
\enann
while then primordial tilt $n_s$ and the tensor-to-scalar ratio $r$ are 
\beann
n_s -1&\simeq & -\frac{8(3 -2\xi \phi^2)}{\phi^2(1-\xi \phi^2)} \left(1+\frac{\lambda \phi^4}{4 M^2(1-\xi \phi^2)}\right)^{-1}
\nonumber \\
&&
~
-2\frac{\lambda \phi^2 (2 -\xi \phi^2)}{M^2 (1-\xi \phi^2)^2} \left(1+\frac{\lambda \phi^4}{4 M^2(1-\xi \phi^2)}\right)^{-2}
\label{gns}
\enann
\beann
r &\simeq &\frac{128 }{\phi^2(1-\xi \phi^2)}\left(1+\frac{\lambda \phi^4}{4 M^2(1-\xi \phi^2)}\right)^{-1}.
\label{gr}
\enann
The e-folding $N$ is given by solving Eqs. (\ref{fr1}),(\ref{fr2}) and (\ref{eom}).
Based on these formula, 
we show the accuracies of the cosmological perturbations in Table \ref{error}.

\vskip 4cm



\end{document}